\documentclass[twocolumn,preprintnumbers,
endnote,nofootinbib,prl]{revtex4}
\usepackage{graphicx}

\newcommand{\vev}[1]{\langle {#1} \rangle}
\newcommand{\lsim}{\lesssim}

\newcommand{\eq}[1]{Eq.~(\ref{#1})}

\newcommand{\beq}{\begin{equation}}
\newcommand{\eeq}{\end{equation}}
\newcommand{\bea}{\begin{eqnarray}}
\newcommand{\eea}{\end{eqnarray}}
\newcommand{\eps}{\varepsilon}

\newcommand{\mzd}{m_{Z_d}}

\newcommand{\sstwmZ}{\sin^2\theta_W(m_Z)_{\rm \overline{MS}}}
\newcommand{\sstwQ}{\sin^2\theta_W(Q^2)}

\begin{document}

% page numbers bottom-center
\pagestyle{plain}

\title{\boldmath Parity Violation and Rare Higgs Decays from a Dark Force}

\author{Hooman Davoudiasl
\footnote{email: hooman@bnl.gov}
}

\affiliation{Department of Physics, Brookhaven National Laboratory,
Upton, NY 11973, USA}

%%%%%%%%%%%%%%%%%%%%%%%%%%%%%%%%%%%%%%%%%%%%%%%%%%%%%%%%%%%%%%%%%%%%%%%%%%%%

\begin{abstract}

We outline the phenomenology of the ``dark" $Z$, denoted by $Z_d$, which is a generalization of the ``dark" photon hypothesis.  
Whereas the dark photon interacts with the Standard Model through kinetic mixing, $Z_d$ is assumed also to have mass-mixing with the $Z$ boson.  In particular, 
we highlight the possibility of $Z_d$ contributions to low $Q^2$ parity violation measurements and rare Higgs decays $H\to Z Z_d \to 4 \ell$, where $\ell$ is a 
charged lepton.  The parity violation effects of a $Z_d$ with an intermediate mass $\sim 10-35$~GeV can in principle relieve the mild $\sim 1.8\sigma$ 
tension among various measurements of the weak mixing angle $\theta_W$.  We briefly comment on the prospects for future parity violation experiments 
at low $Q^2$ to probe this scenario, which could have correlated signals in rare Higgs decays at the LHC.%
\footnote{This article is mainly based on an invited 
talk of the same title presented by the author at  the 22$^{\text nd}$ International Spin Symposium, hosted by the 
University of Illinois and Indiana University, September 25-30, 2016 at University of Illinois Urbana-Champaign.  The topic of that presentation was largely 
the subject of the investigations in Ref.~\cite{Davoudiasl:2015bua}.}

\end{abstract} \maketitle
%%%%%%%%%%%%%%%%%%%%%%%%%%%%%%%%%%%%%%%%%%%%%%%%%%%%%%%%%%%%%%%%%%%%%%%%%

Only about $5\%$ of the energy budget of the Universe is in the form of the ordinary matter made of atoms.  Nearly $27\%$ of that energy budget is 
made up of some unknown substance referred to as dark matter (DM) \cite{Ade:2015xua} and the rest is a type of energy density that 
is responsible for the accelerated expansion of the Universe, often called dark energy.  While dark energy could be the cosmological constant, its nature 
remains as yet uncertain.  

All the evidence that we so far have for DM comes from its gravitational effects and is gleaned from 
various astronomical and cosmological observations.  What we know about DM, broadly speaking, is quite basic: it is a from of matter that is stable on cosmological 
time scales and does not interact with ordinary matter with significant strength if at all.  This leaves a wide range of possible scenarios for what DM could be and 
how it was produced in the early Universe.  The strong evidence for DM is one of the main motivations 
to expect new physics, since none of the known particles within the Standard Model (SM) has the requisite properties to be a viable candidate.  

Many models of DM are based on a single particle that resides in some extension of the SM.  However, it is also possible that DM is part of a ``dark sector" 
that has its own matter content and forces, in close analogy with the SM (which comprises the sub-dominant sector in the cosmic energy budget).  
Such a dark sector could then interact with the SM indirectly, for example through suppressed mixing.  

In what follows, we will assume that the dark sector contains a vector boson $Z_d$ that is the mediator of a 
dark $U(1)_d$ force.  Such a vector boson has been invoked in connection with a potential resolution 
of the $\sim 3.5 \sigma$ deviation in the measured value of the muon anomalous magnetic 
moment $g_\mu - 2$ \cite{Fayet:2007ua,Pospelov:2008zw} 
and also in interpreting certain astrophysical data as signals of DM \cite{ArkaniHamed:2008qn}.  

A popular framework for the interaction of the dark vector boson $Z_d$ with the SM is kinetic mixing \cite{Holdom:1985ag} between 
$U(1)_d$ and hypercharge $U(1)_Y$ through the dim-4 operator
\beq
\frac{1}{2} \frac{\eps}{\cos\theta_W} B_{\mu\nu} Z_d^{\mu\nu}, 
\label{kmix}
\eeq
where $\eps$ is the mixing parameter (assuemd to be small), 
$B_{\mu\nu} \equiv \partial_\mu B_\nu - \partial_\nu B_\mu$, with $B_\mu$ the hypercharge gauge field, 
$Z_{d \mu\nu} \equiv \partial_\mu {Z_d}_\nu - \partial_\nu {Z_d}_\mu$, and $\theta_W$ is the weak mixing angle.  With 
the above coupling, after proper diagonalization of the Lagrangian, one finds that $Z_d$ couples to the 
electromagnetic current $J_{em}^\mu$ via 
\beq
-e \eps J_{em}^\mu {Z_d}_\mu, 
\label{ZdJem}
\eeq
that is like a photon, but suppressed 
by $\eps$.  In this case, the $Z_d$ is often referred to as the ``dark" photon.

The above fairly simple 
setup can in principle lead to an explanation of the $g_\mu -2$ deviation, if the mass of $Z_d$ is 
in the range $\mzd \lsim 1$~GeV and $\eps \lsim 10^{-2}$.  This possibility has provided a well-motivated 
search target for various experiments to look for the potential signals of $Z_d$ in low energy experiments.  If the 
lightest decay final states for $Z_d$ are in the SM sector, with significant branching fractions into $e^+e^-$ and  
$\mu^+\mu^-$, the dark photon explanation of $g_\mu-2$ is basically ruled out by various null search results; see for example 
Ref.~\cite{Alekhin:2015byh}.  

If the dark sector includes states that are lighter than $\mzd/2$ and have non zero $U(1)_d$ charge, then it is typically expected that $Z_d$ would decay dominantly into 
those states and therefore end up being largely ``invisible."  In that case, other constraints will be applicable, which will also 
remove much of the allowed parameter space for explaining the $g_\mu-2$ anomaly; see for example Refs.~\cite{Essig:2013lka,Davoudiasl:2014kua}.  (Since the presentation of 
the original talk, the ``invisible'' dark photon 
explanation of the $g_\mu-2$ anomaly has been essentially ruled out, for $\mzd\leq 8$~GeV, by the BaBar collaboration \cite{Lees:2017lec}.)

Regardless of its utility in addressing the $g_\mu-2$ anomaly, an interesting implication of the above invisibly decaying $Z_d$ scenario is the possibility to produce and detect light dark sector states in fixed target (or beam dump) experiments.  Here, 
the $Z_d$ bosons produced in the collisions of a proton \cite{Batell:2009di} or electron \cite{Izaguirre:2013uxa} 
beam with the target would decay in flight into dark sector states (perhaps DM).  These particles 
would form a ``dark beam," due to the typically large boost of the initial $Z_d$ bosons, and go through a shield (such as Earth) unimpeded.  A suitable detector 
down stream of the beam can then detect the dark states, via the $Z_d$ couplings to the SM ({\it e.g.} kinetic mixing in the case of a dark photon $Z_d$).

One could extend the above picture of $Z_d$ as a ``dark" photon by including $Z_d$-$Z$ mass mixing 
effects parameterized by 
\beq
\eps_Z = (\mzd/m_Z)\delta, 
\label{epsZ}
\eeq
as may be realized in a 2-Higgs doublet model 
with a doublet charged under $U(1)_d$ \cite{Davoudiasl:2012ag}.  With this addition, the couplings of  
$Z_d$ to the SM are given by \cite{Davoudiasl:2012ag}
\beq
{\cal L}_{\text{int}} = (- e \eps J^{em }_\mu - \frac{g}{2 \cos \theta_W} \eps_Z J^{NC}_\mu + \ldots ) Z_d^\mu,
\label{Zd-int}
\eeq
where $J_\mu^{NC}$ is the SM neutral current.  Given the new 
$Z$-like coupling to the neutral current, we refer to $Z_d$ as ``dark" $Z$.  In Eq.~(\ref{Zd-int}), 
$\ldots$ denotes other $Z_d$ couplings, such as $HZZ_d$ \cite{Davoudiasl:2012ag}.  

One can show \cite{Davoudiasl:2012ag} that the above new interactions lead to effects in parity violating processes, as well as rare decays of 
mesons and the Higgs, such as $H\to Z Z_d$, where $Z_d$ could be light and hence on-shell.  For a light dark $Z$, its longitudinal mode 
will dominate the rare decay processes, as the corresponding amplitudes are enhanced by $\sim E/\mzd$, where $E$ 
is the typical energy of $Z_d$ in the process.  Flavor physics data, from $K$ and $B$ decays imply $|\delta| \lsim 10^{-3}$, 
assuming on-shell $Z_d$ final states, depending on  its``visible" branching fractions.  

Assuming both kinetic and mass mixing induced couplings, it can be shown that the contribution of $Z_d$ 
to neutral current amplitudes induces a variation of the weak mixing angle $\theta_W$ with momentum transfer $Q^2$, given by 
\beq
\Delta \sin^2\theta_W (Q^2) = -\eps \delta \left(\frac{m_Z}{\mzd}\right) \frac{\cos \theta_W \sin \theta_W}{1  + Q^2/\mzd^2}\,.
\label{delsintw}
\eeq
Th above shift could yield measurable deviations in the value of the mixing angle for $Q^2\lsim \mzd^2$.

There is currently a slight tension, at $\sim 1.8 \sigma$, between the SM prediction and the measured  
weak mixing angle at various $Q^2 \ll m_Z^2$.  These measurements, extrapolated to the $\overline{\rm MS}$ scale $\mu=m_Z$, are:
\beq
\sstwmZ = 0.2283(20)\quad \text{(APV)}
\label{sin2-APV}
\eeq
\beq
\sstwmZ = 0.2329(13)\quad \text{(Moller E158)}
\label{sin2-E158}
\eeq
\beq
\sstwmZ = 0.2356(16)\quad \text{(NuTeV)}
\label{sin2-NuTeV}
\eeq
at $\vev{Q} = 2.4 ~\text{MeV}$, $\vev{Q} = 160 ~\text{MeV}$, and $\vev{Q} \approx 5 ~\text{GeV}$, respectively.   
The deviation from the SM prediction 
\beq
\sstwmZ = 0.23124(12) \quad \text{(SM)}
\label{sin2-SM}
\eeq
from the weighted average of the above values is 
given by (see Ref.~\cite{Davoudiasl:2015bua} for details): 
\beq
\Delta \sin^2\theta_W \simeq 0.0016(9)\,.
\label{sin2-dev}
\eeq  
%%%%%%%%%%%%%%%%%%%%%%%%%%
\begin{figure*}[t]
\includegraphics[width=0.8\textwidth]{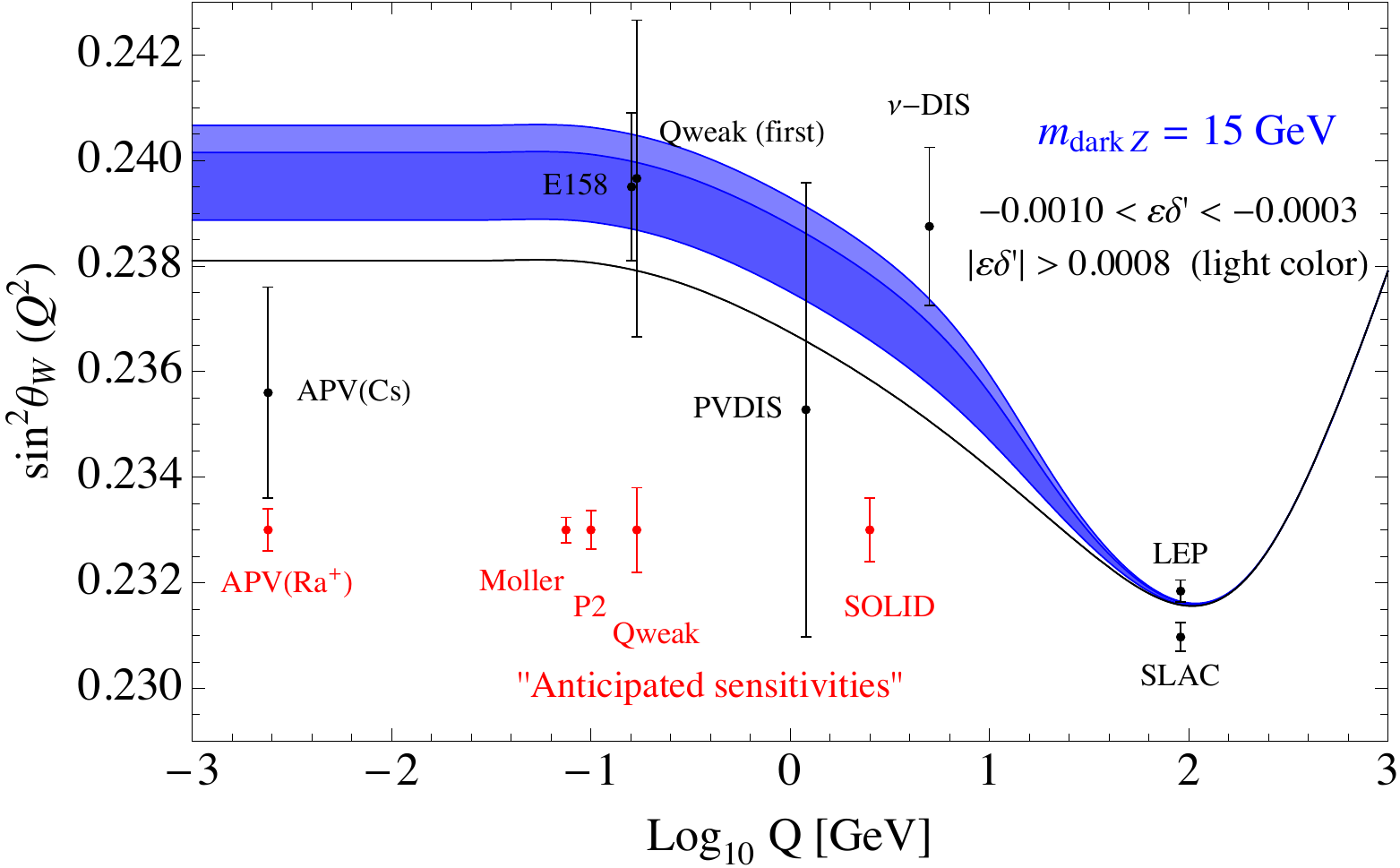}
\caption{Running of $\sstwQ$ with $Q^2$ due to a $Z_d$ with $m_{Z_d} = 15$ GeV for a 1-sigma fit to $\eps \delta'$ from \eq{1sigma}, shown here as the blue band.
The lighter part of the band corresponds to parameters that are in some tension with precision constraints indicated by \eq{epsdelta'} (see text for details).  The black 
curve represents the running of $\sstwQ$ in the SM \cite{Marciano:1980be,Czarnecki:1995fw,Czarnecki:2000ic,Ferroglia:2003wa}.  The sizes of the red 
error bars only represent the anticipated sensitivities of various experiments.  This figure 
is taken from Ref.~\cite{Davoudiasl:2015bua}.
}
\label{fig:sin2twQ2}
\end{figure*}
%%%%%%%%%%%%%%%%%%%%%%%%%%

It is interesting to consider how the contribution of an intermediate mass $Z_d$, with $\mzd \sim 10-35$~GeV, may help 
bring these measurements into better agreement with the SM prediction.  For these values of $\mzd$, the mass mixing parameter 
$\delta$ also receives a non-negligible contribution from kinetic mixing, and we have $\delta \to \delta'$ with 
\beq
\delta' \simeq \delta + \frac{\mzd}{m_Z} \,\eps \, \tan\theta_W\,.
\label{delta'}
\eeq
As an example for illustrative purposes, using the above mass-mixing parameter and setting $\mzd = 15$~GeV, a 
1-$\sigma$ fit to \eq{sin2-dev} would require \cite{Davoudiasl:2015bua} 
\beq
-0.0010 < \eps \delta' < -0.0003\,,
\label{1sigma}
\eeq
as presented in Fig.~\ref{fig:sin2twQ2}.   
This range of parameters is largely 
allowed by the existing constraints on $\eps$ and $\delta'$, as discussed below. 

Constraints from the ATLAS collaboration search 
for $H\to Z Z_d\to l_1^+l_1^-  l_2^+ l_2^-$, with $l_{1,2} = e,\mu$ \cite {Aad:2015sva}, place a 2-$\sigma$ bound on mass 
mixing 
\beq
|\delta'| \lsim 0.02 \quad\text{(ATLAS})\,, 
\label{ATLAS}
\eeq
roughly constant over the $\mzd$ range considered here, assuming that the total branching fraction 
\beq
\text{Br}(Z_d\to l^+l^-) \approx 0.15,   
\label{BrZdll}
\eeq
with $l=e,\mu$.  The bound $|\eps|\lsim 0.04$ is obtained from precision and production constraints 
\cite{Davoudiasl:2015bua} (allowing some cancellation with $\delta'$-dependent contribution).  These constraints suggest the bound
\beq
|\eps \delta'| \lsim 0.0008.
\label{epsdelta'}
\eeq
Hence, a narrow band (light blue in Fig.~\ref{fig:sin2twQ2}) 
in the lower end of the 1-$\sigma$ range (\ref{1sigma}) considered in the above example has some tension with the above bound.  

In the near future, the full results of the Qweak experiment at JLAB are expected to yield a low $Q^2$ determination of $\sin^2\theta_W(Q^2)$ 
with an uncertainty at the $\pm 0.0007$ level.  This would lower the uncertainty of the aforementioned weighted average to $\pm 0.00055$, which 
assuming the central value based on the published results, could lead to a $\sim 3\sigma$ deviation.  At the same time, future LHC search results for the 
rare Higgs decay $H\to Z Z_d$ could probe the relevant parameter space suggested by such a deviation in $\sin^2\theta_W(Q^2)$.  Further reductions in the uncertainty 
of the weighted average of $\sstwmZ$ at low $Q^2$ are expected to be achieved by future measurements of polarized $ee$ Moller scattering at JLAB, as well as polarized 
$ep$ scattering (P2) at MESA, in Mainz.  

In conclusion, we see that the dark $Z$ framework presented here could in principle lead to an interesting 
interplay between various low $Q^2$ effects and high energy LHC signals, originating from 
the contributions of an intermediate mass $Z_d$.

%%%%%%%%%%%%%%%%%%%%%%%%%%%%%%%%%%%%%%%%%%%%%%%%%%%%%%%%%%%%%%%%%%%%%%%%%%%
\acknowledgments

We thank the organizers of the 2016 Spin Symposium for their invitation to present the talk on which this summary was based.  
The work of the author is supported by the United States Department of Energy under Grant Contract DE-SC0012704.

%%%%%%%%%%%%%%%%%%%%%%%%%%%%%%%%%%%%%%%%%%%%%%%%%%%%%%%%%%%%%%%%%%%%%%%%%%%

%%%%%%%%%%%%%%%%%%%%%%%%%%%%%%%%
%%%%%%%%%%%%%%%%%%%%%%%%%%%%%%%%

\bibliography{spin16}

\end{document}